\documentclass[modern]{aastex631}
\usepackage{amsmath}

\shorttitle{Princess Bride Syndrome}
\shortauthors{Kipping}
\graphicspath{{./}{figures/}}

\begin{document}

\title{Exoplaneteers Keep Calling Plots ``Allan Variance'' Plots, \\When They Aren't}

\author[0000-0002-4365-7366]{David Kipping}
\affiliation{Dept of Astronomy, Columbia University, 550 W 120th Street, New York, NY 10027}



\begin{abstract}

I highlight that there is a substantial number of papers (at least 11 published since 2024) which all refer to a specific type of plot as an ``Allan variance'' plot, when in fact they seem to be plotting the standard deviation of the residuals versus bin size. The Allan variance quantifies the stability of a time series by calculating the average squared difference between successive time-averaged segments over a specified interval; it is not equivalent to the standard deviation. This misattribution seems particularly prolific in the exoplanet transit spectroscopy community. However, I emphasize that it does not impact the scientific analyses presented in those works. I discuss where this confusion seems to stem from and encourage the community to ensure statistical measures are named correctly to avoid confusion.

\end{abstract}



\section*{}

\begin{figure}[b]
\begin{center}
\includegraphics[width=12.0cm,angle=0,clip=true]{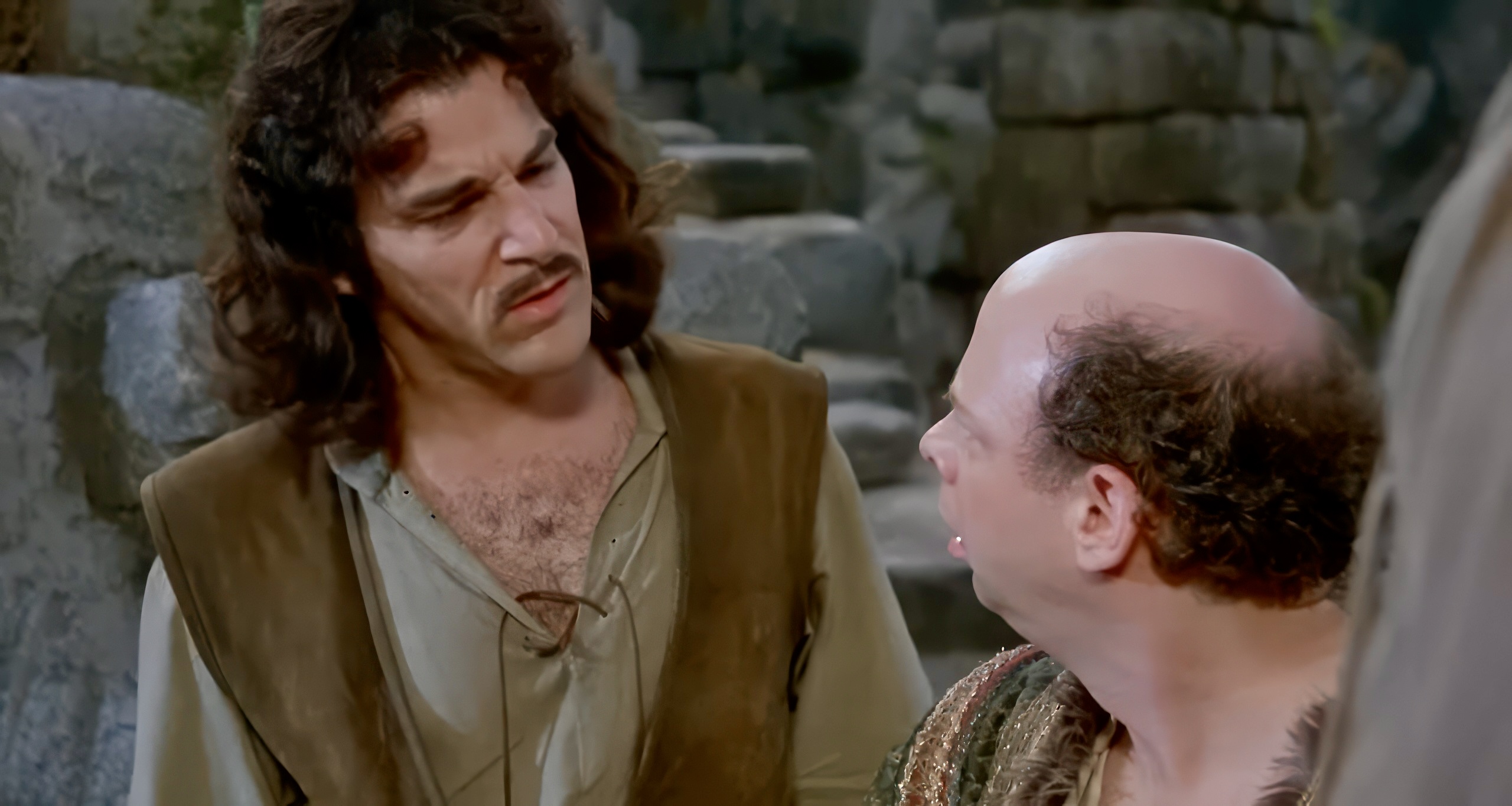}
\caption{\textit{
``You keep using that word. I do not think it means what you think it means.'' Inigo Montoya, Princess Bride
}}
\label{fig:ingio}
\end{center}
\end{figure}

\section{What is the Allan Variance?}

The Allan variance, sometimes dubbed AVAR or the two-sample variance, was first introduced in 1966 by David Allan \citep{allan:1966}. As is often the case, the eponym ``Allan variance'' does not appear in that first publication but was introduced subsequently - in fact by Allan himself four years later \citep{allan:1972}. The concept stems from the study of crystal oscillator stability, in particular in the analysis of ``flicker noise (spectrum proportional to $|\omega|^{-1}$) modulation of the frequency of an oscillator'' \citep{allan:1966}.

Allan sought to quantify the stability of an oscillator by taking a group of frequency measurements over some interval $\tau$, computing their mean, and then comparing it to the mean of the measurements taken over the subsequent time interval $\tau$. Accordingly, the Allan (=two-sample) variance may be expressed as

\begin{align}
\sigma_A^2(\tau) = \frac{1}{2} \left\langle \left(\bar{y}_{i+1} - \bar{y}_i\right)^2 \right\rangle,
\label{eqn:allan2}
\end{align}

where $\bar{y}_i$ is the mean (over interval $\tau$) over the $i^{\mathrm{th}}$ group of frequency measurements, and thus $\bar{y}_{i+1}$ is over the same but over the consecutive group. Because of this consecutive nature, it is often referred to as the ``two-sample variance''; but Allan also introduced a more generalized version that relaxed this assumption, usually referred to as the ``M-sample'' (Allan) variance:

\begin{align}
\sigma_A^2(M, \tau) = \frac{1}{2M^2}\left\langle \left( \sum_{j=0}^{M-1}\left[\bar{y}_{i+j+M} - \bar{y}_{i+j}\right]\right)^2 \right\rangle,
\label{eqn:allanM}
\end{align}

where the summation from $j=0$ to $M-1$ accumulates the differences between sets of non-overlapping averages and the angular brackets indicate an averaging over the index $i$. In the limit of $M=2$, the M-sample variance returns the two-sample variance, although alternative conventions exist which use $M=1$ as the limiting case of equivalence. It is also noted it is sometimes useful to deal with the ``Allan deviation'' instead, which is simply the square root of the two-sample variance:

\begin{align}
\sigma_A(\tau) = \sqrt{\frac{1}{2} \left\langle \left(\bar{y}_{i+1} - \bar{y}_i\right)^2 \right\rangle}.
\label{eqn:allandeviation}
\end{align}

Further, there are also extensions using overlapping intervals and additional averaging steps \citep{allan:1987,rutman:1991}, but the basic concept remains the same.

\section{Valid Applications of the Allan Variance in Astronomy}

To investigate the historical use of the Allan variance in astronomy, I searched NASA ADS using ``collection: astronomy, =body:`allan variance', property:refereed'', returning 398 results as of $11^{\mathrm{th}}$ April 2025. The first instance of the Allan two‐sample variance being used in astronomy appears in \citet{luck:1979} who consider the accuracy of time-keeping systems in astronomy and discuss the use of Equation~(\ref{eqn:allan2}) in measuring the stability of a caesium standard (see their Figure~2). The Allan variance is invoked in related studies published later, such as \citet{ptavcek:1982} and \citet{meinig:1985}.

\begin{figure}
\begin{center}
\includegraphics[width=12.0cm,angle=0,clip=true]{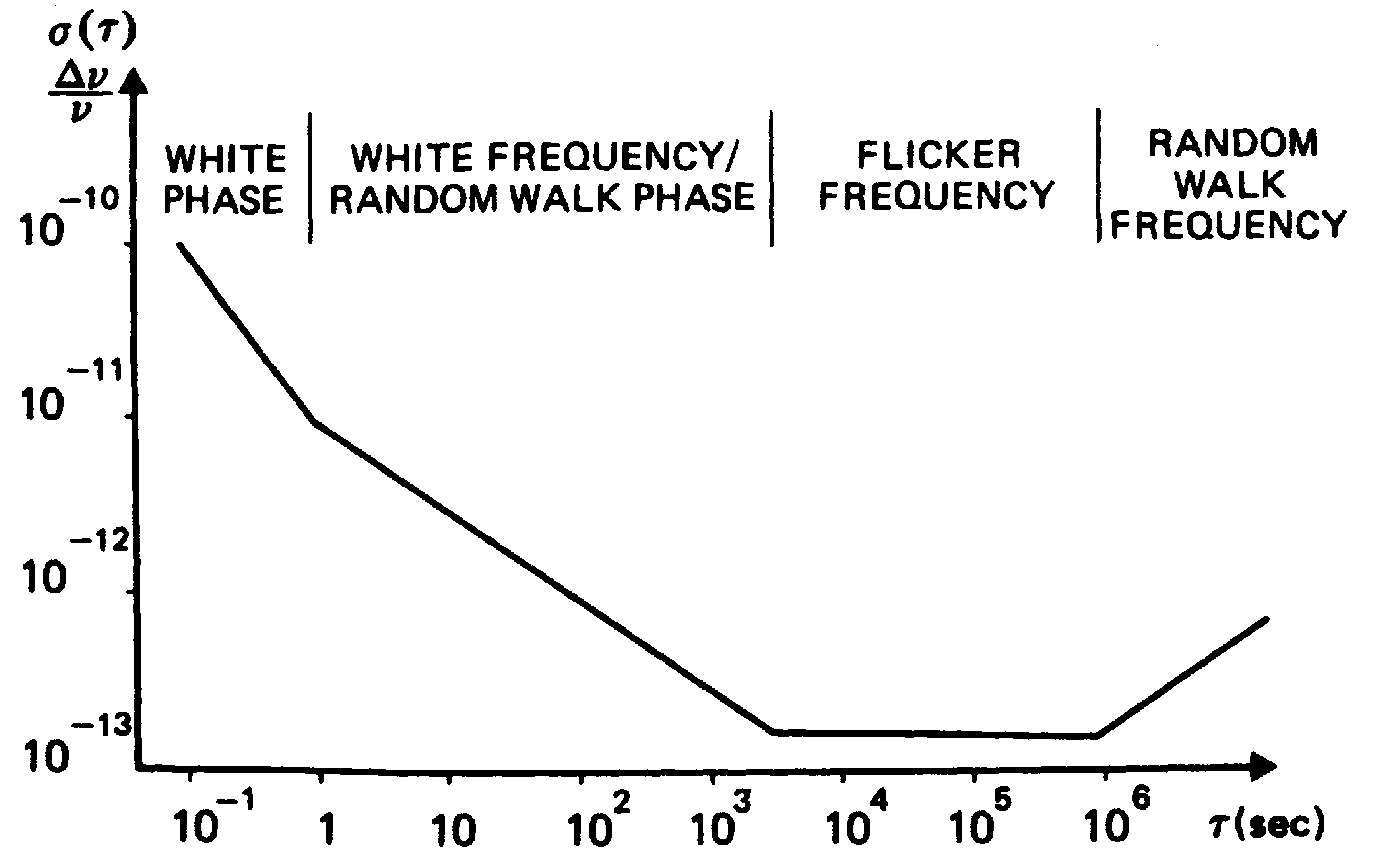}
\caption{\textit{
Reproduction of Figure~2 of \citet{luck:1979}, which appears to be the first application of the Allan variance in a published astronomy article. Although this is only a conceptual figure, it is a correct interpretation of the Allan variance.
}}
\label{fig:ingio}
\end{center}
\end{figure}

Subsequent works continue to focus on the Allan variance as a tool in precise timing experiments, such as spacecraft timing \citep{armstrong:1979,peng:1988,kirsher:1991}, pulsar timing \citep{cordes:1985,guinot:1988,guinot:1991}, testing relativity \citep{vessot:1989,krisher:1990,wolf:1995} and instrumentation \citep{bartholdi:1984,costa:1992,lea:1994}. It is not this author's intention to provide a detailed review of the Allan variance in astronomy, but these examples serve as a typical and appropriate applications of the Allan variance concept, and indeed these fields have continued to use the technique without issue since.

\section{What the Allan Variance Is Not}

In recent years, there are many astronomy papers which appear to incorrectly label a certain type of plot as an ``Allan variance plot'' (see Figure~\ref{fig:ingio}). The misattributed plot corresponds to that of the standard deviation of the residuals versus bin size; hereafter, I refer to this method as the ``time-averaging'' test or plot.

This technique appears to have first been introduced to astronomers in \citet{pont:2006} - a paper that focusses on the impact of red noise on exoplanet transits. In particular, see their Figure~5. The authors note for uncorrelated noise, one expects the standard deviation to vary as $\sigma_1/\sqrt{N}$, where $N$ is the number of points included in each bin. One can write the time-averaging metric of interest as

\begin{align}
\sigma(N) &= \sqrt{ \frac{1}{M} \sum_{i=1}^M \Big( y_i(N) - \overline{\textbf{y}(N)} \Big)^2 },
\label{eqn:timeaveraging}
\end{align}

where $\textbf{y}(N)$ is a vector of the $y$-values binned to a window of $N$ points. It is noted in \citet{pont:2006} that, for independent Gaussian (white) noise, this function has an expectation value of

\begin{align}
\mathrm{E}[\sigma(N)] &= \frac{\sigma_1}{\sqrt{N}},
\label{eqn:Etimeaveraging}
\end{align}

where $\sigma_1$ is the standard deviation of the unbinned $y$-values.

Hopefully, it is clear to the reader that the standard deviation versus bin size (time-averaging test) is not equivalent to the Allan deviation versus lag, by simple inspection of Equation~\ref{eqn:timeaveraging} and Equation~\ref{eqn:allandeviation}. The former is simply a measure of the scatter, whereas the latter is sensitive to autocorrelation. Certainly, both can be useful in testing for red noise \citep{carter:2009}, but they are categorically not the same thing.

At the time of writing, there is widespread incorrect attribution of time-averaging plots as Allan variance plots, particularly in the exoplanet transit spectroscopy community. Such instances are immediately obvious because the $y$-axis is typically labeled noise/RMS/standard deviation, which (as established already) is not equivalent to the Allan deviation or variance. An exhaustive catalog of every misattribution is beyond the scope of this paper, but for the purposes of demonstrating that the problem is indeed widespread, I list 11 papers published since 2024 which mislabel time-averaging test plots as Allan variance plots:

\begin{itemize}
\item Figure~11 of \citet{mukherjee:2025}
\item Figures~2 \& A1 of \citet{kirk:2025}
\item Figure~9 of \citet{bennett:2025}
\item Figure~4 of \citet{gressier:2025}
\item Figure~3 of \citet{rathcke:2025}
\item Extended Data Figure~1 of \citet{murphy:2024}
\item Extended Data Figures~3 \& 4 of \citet{fu:2024}
\item Figures~3, 5 \& 6 of \citet{wallack:2024}
\item Figure~8 of \citet{challener:2024}
\item Figure~2 of \citet{kirk:2024}
\item Figure~C1 of \citet{dyrek:2024}
\end{itemize}

In reference to the above list, through private correspondence, S. Mukherjee confirmed that for Figure~11 of \citet{mukherjee:2025} ``we had used the numpy.std python function with ddof=0 to calculate the standard deviation on the y-axis and the x-axis is the window size used for binning''. Similarly, J. Kirk also confirmed this proceedure.

It is worth noting that none of these cite \citet{allan:1966} or any other papers by Allan. There are several other examples published since 2024 where authors state they inspected their Allan variance plots for indications of red noise, but don't actually show such plots or provide formulae \citep{schlawin:2024,powell:2024,beatty:2024} and thus this author cannot conclusively state they have also made this mistake, although it is possible.

\section{When and How Did This Start?}

\begin{figure}
\centering
\epsscale{1.1}
\plottwo{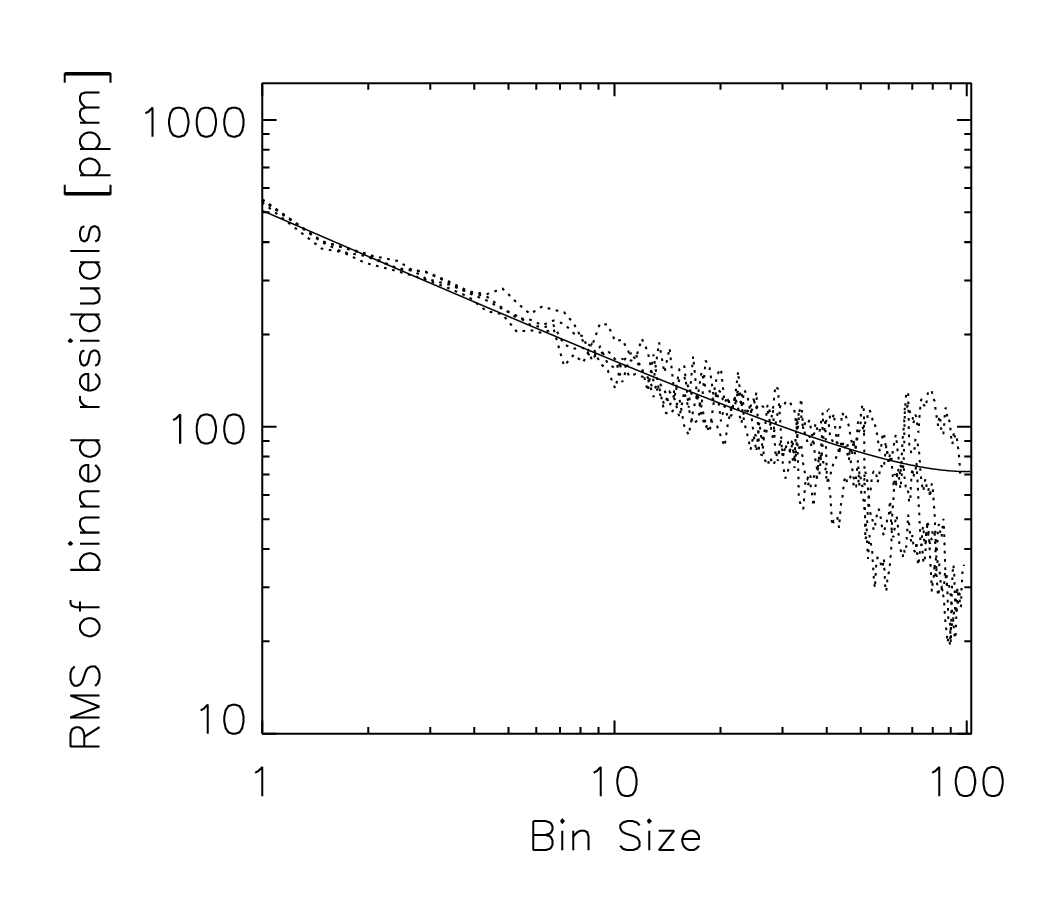}{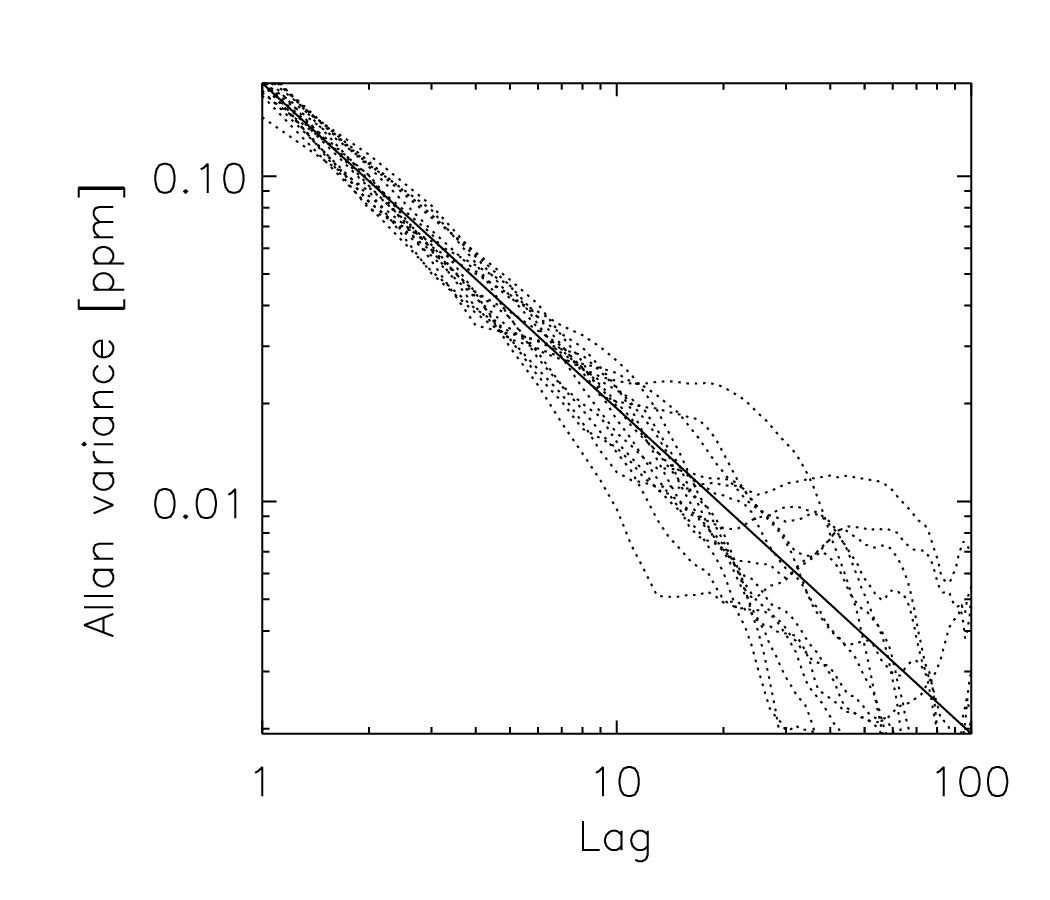}
\caption{\textit{
Reproduction of Figure~4 of \citet{carter:2009}, which appears to be the first published astronomy paper to use the Allan variance to search for time correlated noise in a time series (right-panel). The left-panel is \textbf{not} the Allan variance (but is frequently erroneously called such), but rather it is the RMS versus bin-size of the residuals, also known as the time-averaging test. The use of the Allan (M-sample) variance (right-panel) is innovative, but appropriate and valid.
}}
\label{fig:carter}
\end{figure}

It is fair to describe \citet{pont:2006} as a seminal paper in the study of time-correlated noise in exoplanet time series. This paper introduced the community to the time-averaging test, but it contains no mention of the Allan variance as such a test. The earliest literature instance known to this author appears in \citet{carter:2009}. However, it must be emphasized that \citet{carter:2009} use the Allan variance correctly. Their Figure~4 (reproduced here in Figure~\ref{fig:carter}) provided a direct side-by-side comparison of the time-averaging test popularized by \citet{pont:2006}, as well as a plot of the Allan variance as a function of lag. Again, this comparison further demonstrates that these two metrics are not the same thing (else the two panels would be identical). Nevertheless, the use by \citet{carter:2009} of the Allan variance as a red noise test appears novel in the literature. In their implementation, \citet{carter:2009} use the M-sample Allan variance (Equation~\ref{eqn:allanM}) and note that independent, Gaussian residuals would have an expectation value of

\begin{align}
\mathrm{E}[\sigma_A(\tau)] &= \frac{\sigma_A(0)}{\sqrt{N}}.
\label{eqn:Eallan}
\end{align}

To our knowledge, the next literature instance of the Allan variance in exoplanet studies was by this author, in \citet{kipping:2011}. That paper's implementation is identical to that of \citet{carter:2009}, again showing a side-by-side comparison of the time-averaging test and the Allan M-sample variance in Figure~2 of that work. \citet{clanton:2012} appears to be the next exoplanet-related study using the Allan variance, but again uses it correctly.

The first instance of misuse, at least in the published exoplanet literature, seems to begin with \citet{swift:2015}, where both their Figures~17 and 19 claim to be an ``Allan Variance Plot'' - but both label the $y$-axis as ``RMS (mmag)'', which of course is a distinct quantity from the Allan deviation/variance (e.g. see Figure~\ref{fig:carter}). This paper was well-cited (71 citations to date) and thus could plausibly be the seed for the later snowball of misattributions.

\section{Stigler's Law}

Stigler's Law of Eponymy states that no scientific discovery is named after its original discoverer. Here, it has been demonstrated that Allan variance plots are subject to widespread miseponymy but the intent of this short paper is to try to stop the wave before we completely lose sight of historical accuracy. In particular, the test of plotting RMS versus bin size, popularized by \citet{pont:2006}, is frequently mislabeled as an Allan variance plot - especially in the exoplanet transit spectroscopy community. However, the good news is that such mislabeling does not appear to impact any of the scientific conclusions in those papers. Those authors used the time-averaging test to look for red noise, which is precisely what it's meant for \citep{pont:2006} - it's merely an issue of nomenclature.

Nevertheless, correcting the name is clearly important, not just for historical accuracy but because \textit{there is} a statistical measure called the Allan variance that describes something distinct. There are certainly similarities in both, such as having the same units, being plots of the measure over something like a timescale, both monotonically decrease and even share the same form for how the expectation value drops as the timescale increases (Eqautions~\ref{eqn:Etimeaveraging} \& \ref{eqn:Eallan}). The fact that these metrics were first presented side-by-side in \citet{carter:2009} and later in \citet{kipping:2011} may have contributed to the confusion.

Hopefully, a good lesson from this is to cite the original statistical sources - in this case \citet{allan:1966} - and actually check that the formula one is using is indeed that originally published. Citing by proxy risks a game of telephone, where eventually concepts get distorted - as appears to have happened here.

\vspace{2cm}
Thanks to Ben Cassese and Daniel Yahalomi for conversations which assured me I wasn't totally insane. Also thanks to Sagnick Mukherjee and James Kirk for helping clear up this issue.

%

\vspace{5mm}






\bibliography{manuscript}{}
\bibliographystyle{aasjournal}



\end{document}